\documentclass[10pt,times,twocolumn]{article}
\usepackage{latex8}

\usepackage{ifpdf}
\ifpdf
\fi
\usepackage[T1]{fontenc}

\usepackage{url}
\usepackage{floatflt}
\usepackage[tight]{subfigure}
\usepackage{latexsym}
\usepackage{clrscode}
\usepackage{graphicx}

\author{
Thomas C. Schmidt
\\t.schmidt@ieee.org
\and
Matthias W\"ahlisch\thanks{The author is also with link--lab, H\"onower Str. 35, D--10318 Berlin, Germany.} 
\\waehlisch@ieee.org
\and
Ying Zhang\\ zhang.ying@web.de
\and
HAW Hamburg, Dep. Informatik,
 Berliner Tor 7, D--20099 Hamburg, Germany \\
}
\title{On the Correlation of Geographic and Network Proximity at Internet Edges and its Implications for Mobile Unicast and Multicast Routing.\thanks{This work is supported by the German Bundesministerium f\"ur Bildung und Forschung within the Project {\em Moviecast}. The authors thank CAIDA for making Skitter
datasets available.}
}
\date{}

\begin{document}
\maketitle
\thispagestyle{empty}
\begin{abstract}
Significant effort has been invested recently to accelerate handover operations in a next generation mobile Internet. Corresponding works for developing efficient mobile multicast management are emergent. Both problems simultaneously expose routing complexity between subsequent points of attachment as a characteristic parameter for handover performance in access networks.

As continuous mobility handovers necessarily occur between access routers located in geographic vicinity, this paper investigates on the hypothesis that geographically adjacent edge networks attain a reduced network distances as compared to arbitrary Internet nodes.
We therefore evaluate and analyze edge distance distributions in various regions for clustered IP ranges on their geographic location such as a city. We use traceroute to collect packet forwarding path and round-trip-time of each intermediate node to scan-wise derive an upper bound of the node distances.  Results of different scanning origins are compared to obtain the best estimation of network distance of each pair. Our results are compared with corresponding analysis of CAIDA Skitter data, overall leading to fairly stable, reproducible edge distance distributions. As a first conclusion on expected impact on handover performance measures, our results indicate a general optimum for handover anticipation time in 802.11 networks of 25 ms.
\end{abstract}
\Section{Introduction}

Mobile environments, devices and applications are one of the major driving forces for technological development today, while deployment is still dominated by non--IP appliances. However, the roadmaps of converged services for Next Generation Networks (NGNs) \cite{etsi-180001} on the one hand, efficient mobility management within the next generation Internet \cite{rfc-3775} on the other, lead expectations to the Internet layer as the prevalent tie for mobile access technologies and services.

Seamless support for Voice over IP (VoIP) and related real--time communication must be considered  critical for deployment success into the mobile world. Therefore significant  effort is continuously  taken in the IETF to develop and improve protocols for seamless mobility handovers, FMIPv6 \cite{rfc-4068} and HMIPv6 \cite{rfc-4140} being the most prominent examples. The IP layer introduces scalable multicast as supplementary function, which will be of particular importance to multimedia group conferences in mobile environments of limited capacities. Seamless mobility extensions to IP--layer multicast are likewise under preparation \cite{rfc-draft-mmcastv6-ps-01}.
 
Mobile IPv6 inherits a strong topology dependence through its binding update procedures with the Home Agent (HA) and the Correspondent Node (CN). Handover acceleration schemes attempt to overcome this obstacle by relocating immediate transfer negotiations to the vicinity of the mobile node, i.e., to access networks at the Internet edges. In previous analysis \cite{sw-pvrah-05} it could be shown that the actually attained handover performance largely depends on the relative network topology of access components, when measured in an appropriate delay metric such as round trip time. Access router distance can be considered as the characteristic complexity parameter in fast or hierarchical mobile IPv6. 

Similar observations hold for mobile multicast. While multicast listeners may rely on handover operations derived from  unicast protocols, mobile multicast sources need to reshape distribution trees. Due to the self--similar nature of shortest path trees, alteration requirements of multicast forwarding states are minimally bound by a characteristic function of source displacement.
As shown in \cite{sw-mdtem-06} the routing cost of mobile multicast source management in SPT-based protocols is directly dominated by the topological hop distance attained between designated routers at the previous and next point of attachment.

In this work we empirically analyze the regional edge distance distributions of exemplary areas in the current Internet. We perform selective scans and evaluate the data pool of the CAIDA {\em Skitter} project and derive mobility performance spectra thereof. The paper is organized as follows. We introduce our evaluation methodologies, discuss problems, limitations and related work in the following section. Measurement results are presented and discussed in section \ref{res}. Section \ref{cons} briefly derives consequences for characteristic handover performance measures  of our data. Finally, section \ref{c-o} is dedicated to conclusions and an outlook.

\Section{Methodology and Related Work}
\label{met}

\SubSection{Edge Distance Estimates}
\label{met-ed}

The objective of this work is to inquire on the distribution of network distances between pairs of Internet edge routers located within geographic vicinity. According to the mobility investigations introduced above, network distance measures require metrics of delay and routing complexity, e.g., round trip time and hop count. Scanning  can thus be performed by traceroute and will proceed as follows.

\begin{figure}
  \center
  \includegraphics[scale=0.4]{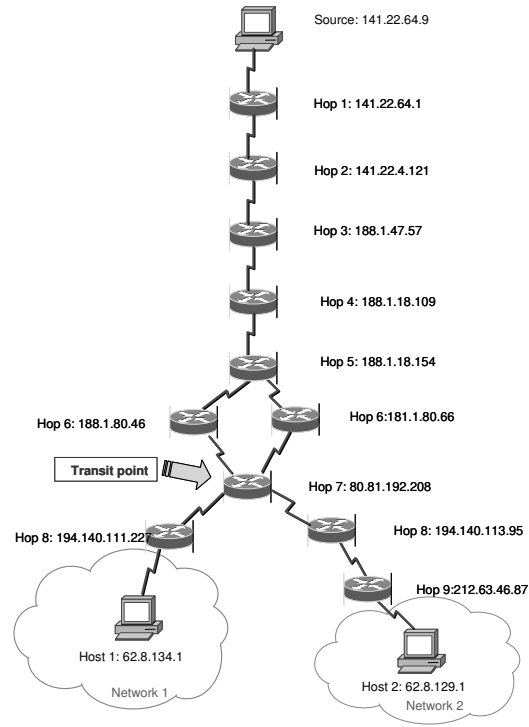}
  \caption{Transit Point Discovery with traceroute}
  \label{fig-intersect}
\end{figure}

\begin{figure}
  \center
     \includegraphics[scale=0.4]{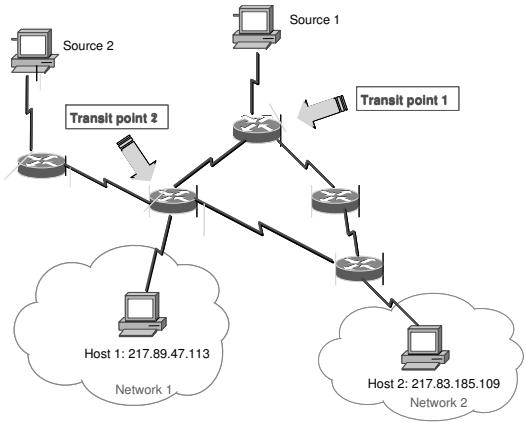}
  \caption{Approximation by Multisource Scanning}
  \label{fig-multintsect}
\end{figure}

Clusters of IP ranges from geographic regions such as cities are pre-selected in order to account for locality. A life host is determined from each range, such that for each pair of hosts the network distance of their corresponding access routers can be evaluated as a sample probe. The appropriate method to determine a routing path between two Internet hosts from a remote location is given by loose source routing. Since source routing is mainly unsupported throughout the Internet, we proceed by auxiliary means as shown in figure \ref{fig-intersect}: Both routing path from the origin to the hosts under consideration are evaluated via traceroute \cite{j-trrt-89} and compared. The last common hop then is determined as a potential transit point, giving rise to an upper bound of the network path between the target nodes via the discovered transit point. Such Evaluation is done under the assumption of symmetric routing (cf. section \ref{met-trrt} on this assumption) and from the perspective of the source. Scan experiments therefore are repeated from multiple origins, eventually discovering closer transit points and shorter neighboring paths as displayed in figure \ref{fig-multintsect}. Results are then derived as minimal upper bound using the path via the closest transit point.

Geographic locations of IP address ranges need provisioning as external input to the scanning. To choose for a reliable source of geographic information we evaluated eight different mapping resources in a first step by selecting a set of 30 distributed, geographically known IP ranges. The commercial product GeoIP \cite{maxmind} thereby was the only resource to admit negligible errors. In a second, automated testing we compared  data of larger samples with {\em whois} queries and found a coincidence rate of about 80 \%. This result we considered reasonable, as {\em whois} data commonly provide administrative addresses possibly distinct of physical router locations.

It should be mentioned that measurements have been undertaken for IPv4 only, even though the initial motivation was derived from mobile IPv6 handovers. Due to our limited accessibility of IPv6 communication this procedure is justified by the expectation that quantitatively dominating dual stack access networks will attain identical topological properties. 

\SubSection{Scanning Problems with traceroute}
\label{met-trrt}

The well known traceroute utility \cite{j-trrt-89} sends out TTL restricted UDP probe packets to inquire on all forwarding nodes along a path to a given destination. It thus leads to a complete topological path vector of the examined route. However, a fair number of nodes throughout the Internet does not reply correctly to standard traceroute queries for several reasons. At first, UDP forwarding may be blocked at some router, in which case traceroute can be switched to use ICMP probes. If the latter are blocked, as well, probing is successfully inhibited. At second, ICMP replies are sometimes suppressed or manipulated by ISPs, which only interferes with our measurements if this happens at or beyond a transit node. The occurrence of routing loops beyond the transit point will likewise lead to discarding the event. The most frequent reason for obtaining invalid data results from asymmetric routes, discoverable by decreasing cumulative round trip times. In a general attempt Paxson \cite{p-erbi-97} analyzed 40,000 end--to--end paths and identified half of them as asymmetric. As our experimental concern concentrates on edge topologies, measurements remain unaffected by asymmetries within the core, leading to significantly lower event rejection rates. Altogether the success ratio in our measurements varies between 45 \% and 65 \%.

\SubSection{Related Work}
\label{met-rw}

Internet topology has been studies for over ten years. These studies focus on characterizing and delineating Internet topology and performance. CAIDA \cite{caida} is a prominent group of pioneers, who record and measure Internet data continuously for almost ten years. The Mercator project \cite{gt-himd-00} uses hop-limited probes in the same manner as traceroute, to infer an Internet map at the route-level. A highly distributed scanning approach is taken within the DIMES project \cite{ss-dlimi-05} with the aim of increased accuracy and comprehensiveness. Aside from many others, Janic and Van Mieghem \cite{jm-opmrt-06} recently performed traceroute scans to investigate the node degree distribution in the Internet and report about complicacies similar to our observations.

Since IP addresses are location-independent, there has been much work on the problem of correlating IP addresses to geographic locations. One of the latest studies was done by Subramanian et al. \cite{spk-gpir-02}. They suggest that 90 \% of nodes within 5 ms RTT are located within a radius of 50 km and 90 \% of nodes within 10 ms RTT are located within a radius of 300 km. 
A CAIDA group \cite{hfmnc-mitar-00}  has studied  network connectivity in the Asia-Pacific region, mainly focusing on network latency and performance,  country peering and third party transit. They use skitter to measure the forward IP path and round-trip-time to about 2,000 destinations in the Asia-Pacific region from different sources and conclude that geographic proximity reflects only in round trip times, while hop count is not a representative metric therefor.

\Section{Regional Edge Distance Distributions}
\label{res}

In this section we present the results for the cities of San Francisco, USA, Berlin and Hamburg, Germany, and Shanghai, China, which were exemplarily selected as geographic target regions. Scanning has been performed from September to December 2006, originating from the locations of Berlin, Hamburg and Shanghai, at the 67th IETF meeting in San Diego and with the help of various public traceroute facilities.\footnote{Public traceroute services did not provide ICMP probing and resultantly provided only little success rates.} Since the number of available IP ranges vary from Shanghai (763) up to San Francisco (8476), subsets of equal sizes are selected randomly for each city. Statistical convergence with respect to sample size, but also for different dates and day times were compared, and a fair stability of the distributions could be observed for sample subsets of 500 IP ranges.

\begin{figure*}[t]
  \center
  \subfigure[Scan Data]{
     \includegraphics[scale=0.65]{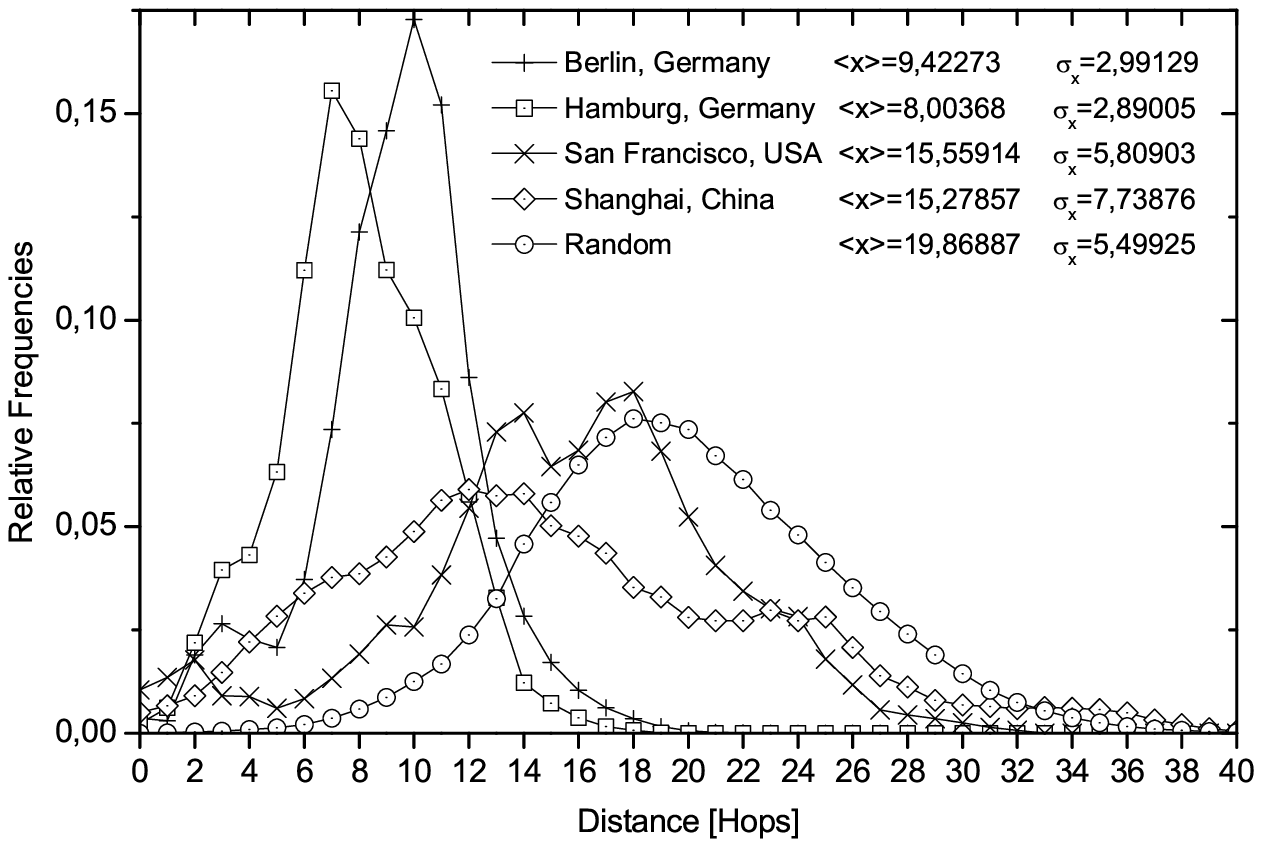}
  }
  \subfigure[CAIDA Data]{
     \includegraphics[scale=0.65]{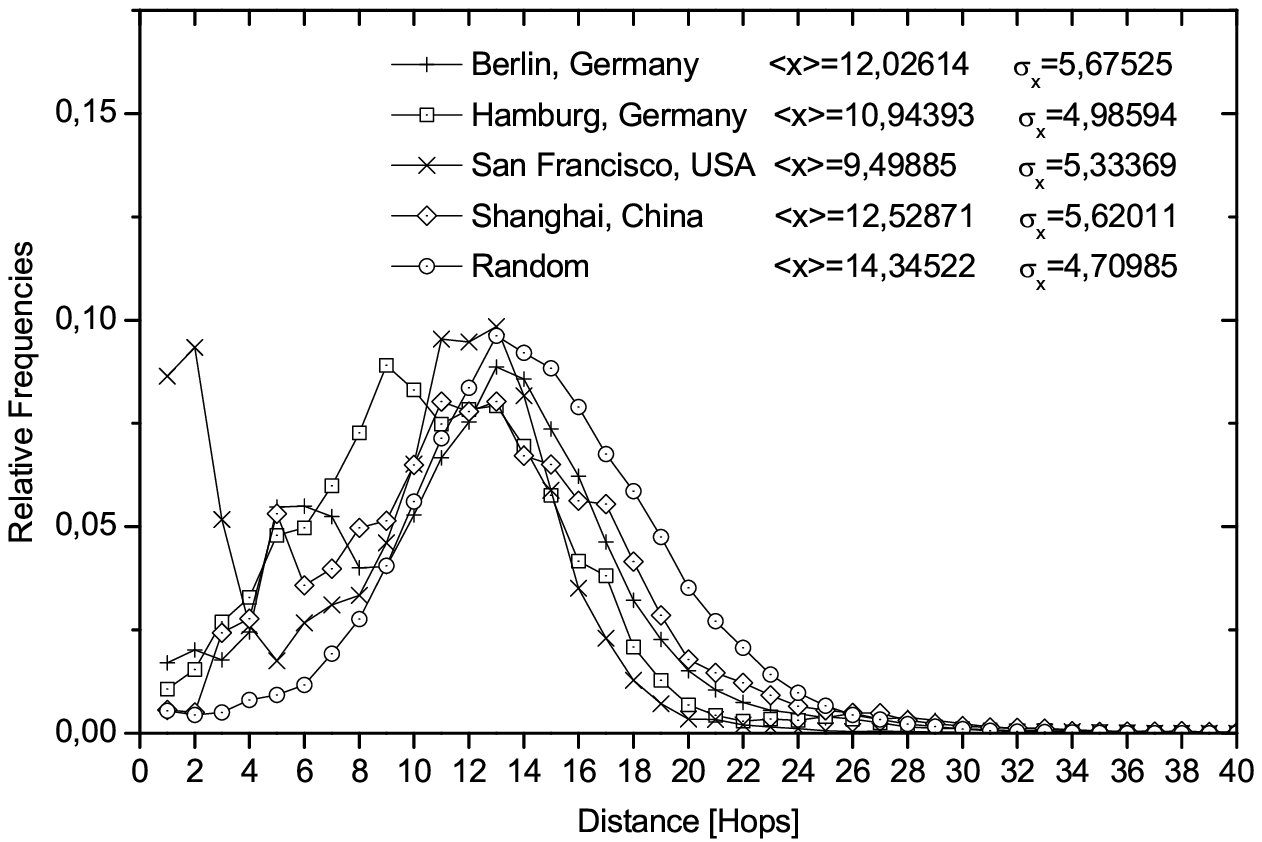}
  }
\caption{Hop Count Distributions at Network Edges in Four Cities. $< x >$ and $\sigma_{x}$ represent mean and standard deviation of the corresponding distributions.}
  \label{fig-hops}
\end{figure*}

\begin{figure*}
  \center
  \subfigure[Scan Data]{
     \includegraphics[scale=0.65]{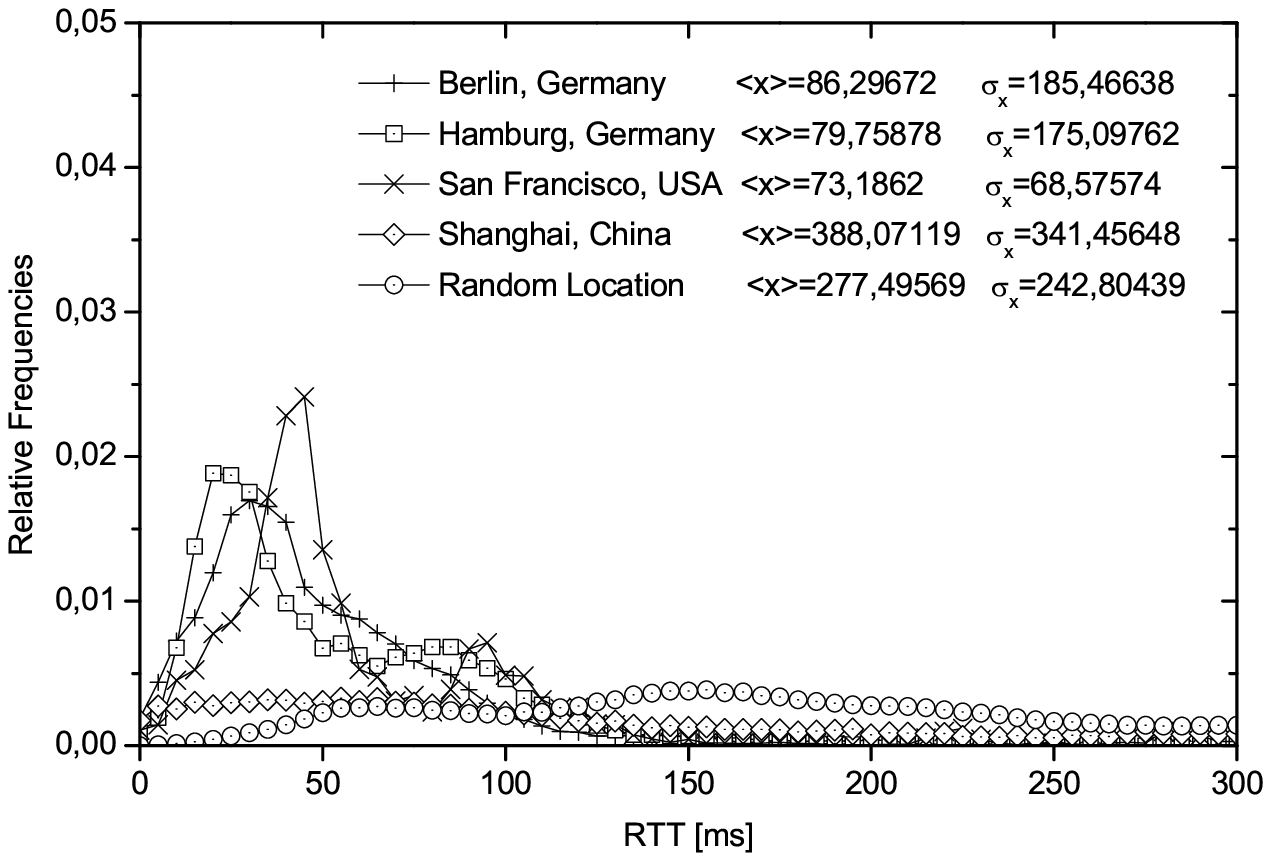}
  }
  \subfigure[CAIDA Data]{
     \includegraphics[scale=0.65]{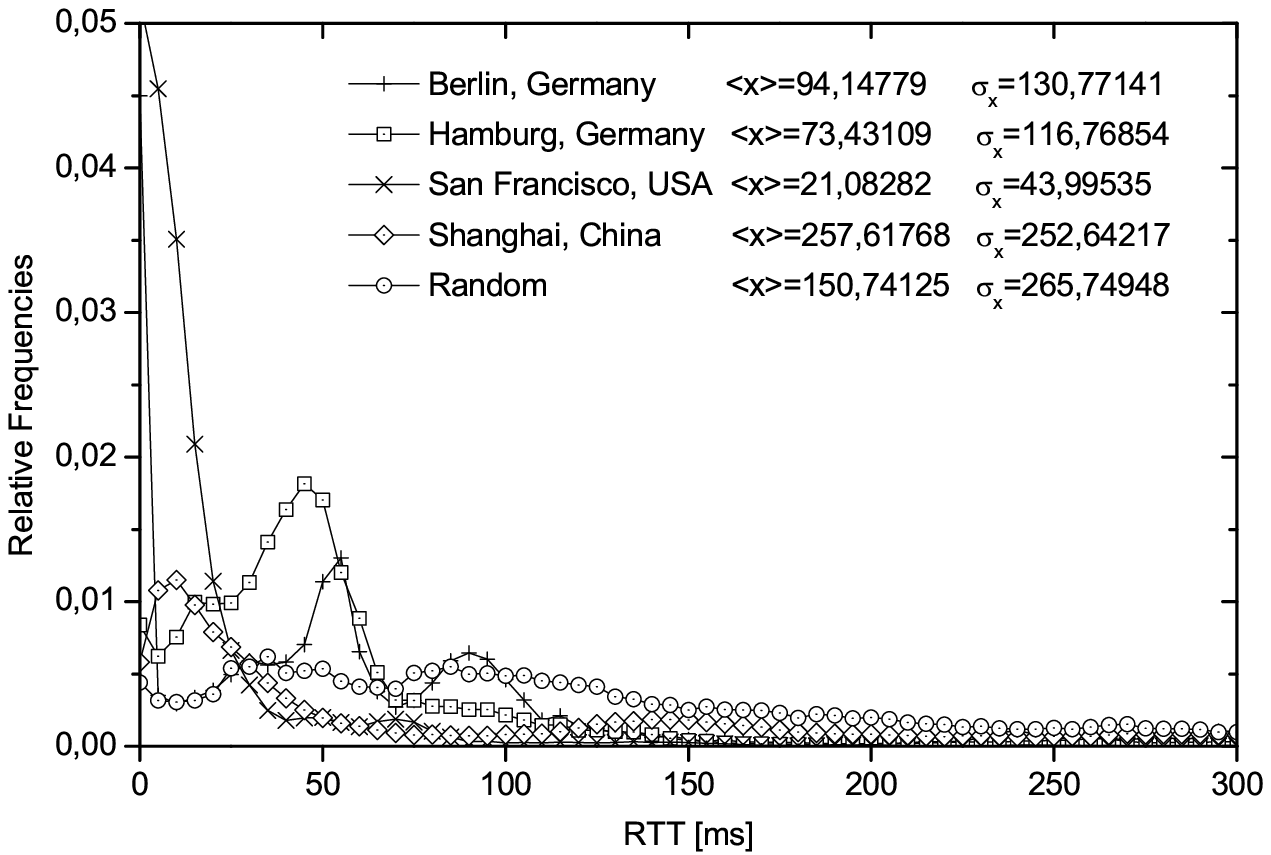}
  }
  \caption{Round Trip Time Distributions at Network Edges}
  \label{fig-rtts}
\end{figure*}

We compare our results with distributions derived from CAIDA data recorded in October 2006. Host clusters for selected cities are taken from the CAIDA destination list according to the GeoIP database. Trace paths are minimized with respect to all available 18 monitor points, which are located more densely at the US West Cost and sparsely in Europe and Asia.

Resulting distributions for hop counts and round trip times are displayed  in figures \ref{fig-hops} and \ref{fig-rtts} respectively. An additional curve derived for randomly located nodes is added to distinguish locality correlations. 

Clearly our hop count results vary significantly, while CAIDA skitter data evaluate to fairly similar distributions.  For Berlin and Hamburg we measure clear peaks at around 8 hops and underestimate those values of CAIDA, while we slightly overestimate hop counts for San Francisco.  These differences may be explained by monitor point positions. While we had direct access to several networks located in Berlin and Hamburg, skitter data are not available from any monitor point close to the two cities or even in Germany. Some CAIDA monitor points are located in close vicinity of San Francisco, e.g., in San Jose, whereas significant parts of our distribution was built from scans at San Diego origin. From this argument it may be concluded that our data better approximate edge distributions of the European cities, while CAIDA values reflect those of San Francisco in a better quality.

\begin{figure}
  \includegraphics[scale=0.65]{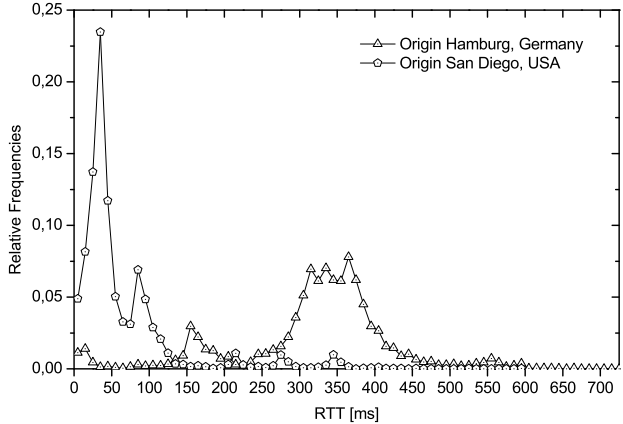}
  \caption{RTT Distributions Evaluated from Different Origins for San Francisco Nodes}
  \label{fig-dior}
\end{figure}

Hop counts in Shanghai admit much wider distributions, even though data are mainly recorded from a source in Shanghai. Scans originating from Europe even show a geographic anti--correlation when compared to the random sample. These results indicate that regional routing topologies are not densely meshed in the Chinese city, such that geographically neighboring access networks are mainly connected via a general degree of indirection. On the contrary, Berlin and Hamburg results expose a pronounced location--to--network hop distance correlation, which somewhat contradicts the conclusions drawn in \cite{hfmnc-mitar-00}. Caida monitors seem to be too sparsely distributed to catch a clear, distinguishable view on hop count laws at Internet edges.

Round trip time distributions exhibit similar behavior. Pronounced peaks at close distances can be observed for the areas of Berlin and Hamburg, when monitored from the close vicinity. The effect of scanning source positioning on RTT results  is shown for San Francisco data in figure \ref{fig-dior}. RTT characteristics, though, appear heavier tailed than hop counts, which supposedly is due to sporadic slow transition links. Tardy transitions are of lesser effect in the San Francisco region, for which again CAIDA measurements segregate a distribution of higher significance.

In contrast, Shanghai data attain merely indifferent RTT distributions, which are even less pronounced than random samples. Non-negligible weights are situated beyond the displayed interval, as can be read from mean and standard deviation values. This may be explained from a wide variety of slow transit links present in the Chinese core networks.\footnote{Another possible explanation could lie in a reduced accuracy of MaxMind GeoIP data for the Chinese region.} 
 CAIDA skitter data seem to qualitatively reflect these RTT law diversities.  San Francisco  values are very pronounced, whereas Hamburg and Berlin data show an intermediate characteristic. It should be noted, though, that the reverse of the proximity observations from \cite{spk-gpir-02} does not seem to hold: RTT distributions admit wide tails, whence even in close router distances enhanced mutual delays may be expected.

In total the results seem to indicate that inter--edge routing within a geographic region is frequently performed via local transits and peering, which produce network proximity in 'the neighborhood', but remain invisible for a distant monitor.

\Section{Applications to Handover Performance}
\label{cons}

The results obtained so far may serve as an empirical fundament for realistic handover performance estimates of the network. A mobile node moving from one access network to another in geographical neighborhood does imposes traffic redirection, minimally from its previous to its new attachment. These operations cause delay and routing costs, which for the case of FMIPv6 \cite{rfc-4068} are given by the unicast path from previous to next access router, and higher, otherwise.

\begin{figure}
  \includegraphics[scale=0.65]{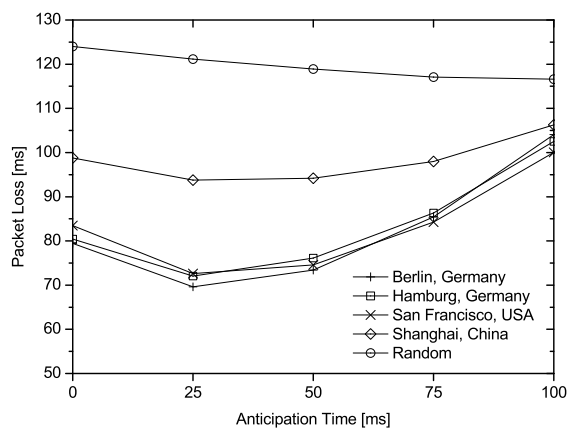}
  \caption{Expected Packet Loss in Predictive Handovers as a Function of Anticipation Time (0 = Reactive Handover)}
  \label{fig-hov}
\end{figure}

Based on the results derived in \cite{sw-pvrah-05} we now can immediately calculate expected values of characteristic handover measures. For packets sent at a constant bit rate of one per 10 ms, the conditional expectation  of packets lost or buffered for given inter--access--router delay was derived for predictive and reactive handover procedures (cf. figure 6 of \cite{sw-pvrah-05}). Combining these previous results with those shown in figure \ref{fig-rtts}, we arrive at expected periods for packet loss as functions of handover anticipation times. Results for the different regions as presented in figure \ref{fig-hov} jointly show a  pronounced uniform minimum at handover anticipation of 25 ms for the cities of San Francisco, Hamburg and Berlin, while significant optimal values remain absent for Shanghai and random data. These results reflect the degree of locality in regional delay distributions. 

Similar conclusions can be drawn for costs needed to reshape shortest path multicast distribution trees under source mobility. In previous simulation studies we derived conditional expectations for multicast forwarding state persistence at a given hop distance between previous and next designated router (cf. figure 3 of \cite{sw-mdtem-06}). Resolving conditioning with the help of above results gives rise to an expected ratio of multicast states persistent under mobility for the regions under consideration. A mobile multicast source feeding a regionally distributed group in San Francisco, Hamburg or Berlin will have to expect on handover the invalidation of about 5 \% of forwarding states at multicast routers, while 10 \% of the routers in a comparable handover situation in Shanghai need to establish new states. Correspondingly, from 75 \% down to 60 \% of multicast states survive a source handover in widely distributed groups.

\Section{Conclusions and Outlook}
\label{c-o}

Quality of service characteristics of the Internet are likely to drive or delay the deployment of all--IP mobile devices. Inspired by mobile IP handover performance measures we analyzed routing distances in geographically bound clusters of the Internet, in which users are expected to move around freely while continuously 'talking' IP in the near future. Traceroute probes have been used to derive hop count and delay distributions at Internet edges in San Francisco, Berlin, Hamburg and Shanghai. Comparison has been drawn to CAIDA measurements. Our results seem to indicate a clear signatures of locality in both distance metrics, which cannot be segregated from CAIDA measurements due to sparsely scattered monitor points. The application of these results to calculating packet loss after mobility handovers indicates that characteristic proximity measures in the Internet may give rise to fairly stable anticipation timers.

In future work we will refine our measures by targeting additional geographic regions and adding PlanetLab nodes to our monitors. We intend to include DIMES data into our comparison, continue to derive QoS distributions and characteristic values from regional delay distributions to put expected handover characteristics of a future mobile Internet on firmer grounds. 

\bibliographystyle{latex8}
\bibliography{fhtw-ipv6,fhtw-vcoip,vcoip,mipv6,mcast,mmcast,mmodeling,ssm,imeasurement}

\begin{thebibliography}{10}\setlength{\itemsep}{-1ex}\small

\bibitem{gt-himd-00}
R.~Govindan and H.~Tangmunarunkit.
\newblock Heuristics for internet map discovery.
\newblock In {\em Proceedings {IEEE INFOCOM} 2000}, volume~3, pages 1371--1380,
  Piscataway, NJ, USA, March 2000. IEEE Press.

\bibitem{hfmnc-mitar-00}
B.~Huffaker, M.~Fomenkov, D.~Moore, E.~Nemeth, and k~claffy.
\newblock {Measurements of the Internet topology in the Asia-Pacific Region}.
\newblock In {\em INET 2000 Proceedings}, Yokohama, Japan, June 2000. Internet
  Society.

\bibitem{j-trrt-89}
V.~Jacobson.
\newblock Traceroute tool.
\newblock ftp://ftp.ee.lbl.gov/traceroute.tar.gz, 1989.

\bibitem{jm-opmrt-06}
M.~Janic and P.~{Van Mieghem}.
\newblock On properties of multicast routing trees.
\newblock {\em Int. J. Commun. Syst.}, 19(1):95--114, 2006.

\bibitem{rfc-3775}
D.~B. Johnson, C.~Perkins, and J.~Arkko.
\newblock Mobility {S}upport in {IP}v6.
\newblock RFC 3775, IETF, June 2004.

\bibitem{rfc-4068}
R.~Koodli.
\newblock Fast {H}andovers for {M}obile {IP}v6.
\newblock RFC 4068, IETF, July 2005.

\bibitem{maxmind}
{MaxMind LLC}.
\newblock {GeoIP}.
\newblock http://www.maxmind.com, 2006.

\bibitem{p-erbi-97}
V.~Paxson.
\newblock {End-to-End Routing Behavior in the Internet}.
\newblock {\em IEEE/ACM Trans. Netw.}, 5(5):601--615, 1997.

\bibitem{sw-pvrah-05}
T.~C. Schmidt and M.~W{\"a}hlisch.
\newblock Predictive versus {R}eactive -- {A}nalysis of {H}andover
  {P}erformance and {I}ts {I}mplications on {IP}v6 and {M}ulticast {M}obility.
\newblock {\em Telecommunication Systems}, 30(1--3):123--142, November 2005.

\bibitem{sw-mdtem-06}
T.~C. Schmidt and M.~W{\"a}hlisch.
\newblock {Morphing Distribution Trees -- On the Evolution of Multicast States
  under Mobility and an Adaptive Routing Scheme for Mobile SSM Sources}.
\newblock {\em Telecommunication Systems}, October 2006.
\newblock in print, appeared online.

\bibitem{rfc-draft-mmcastv6-ps-01}
T.~C. Schmidt and M.~W{\"a}hlisch.
\newblock {M}ulticast {M}obility in {MIPv6:} {P}roblem {S}tatement.
\newblock IRTF Internet Draft -- work in progress~01, MobOpts, October 2006.

\bibitem{ss-dlimi-05}
Y.~Shavitt and E.~Shir.
\newblock {DIMES: Let the Internet Measure Itself}.
\newblock Technical Report cs.NI/0506099, arXiv.org, June 2005.

\bibitem{caida}
{Skitter Project at CAIDA}.
\newblock http://www.caida.org/tools/measurement/skitter.

\bibitem{rfc-4140}
H.~Soliman, C.~Castelluccia, K.~Malki, and L.~Bellier.
\newblock Hierarchical {M}obile {IP}v6 mobility management ({HMIP}v6).
\newblock RFC 4140, IETF, August 2005.

\bibitem{spk-gpir-02}
L.~Subramanian, V.~Padmanabhan, and R.~Katz.
\newblock {Geographic Properties of Internet Routing}.
\newblock In {\em Proceedings of the 2002 USENIX Annual Technical Conference},
  pages 243--259, Berkeley, CA, USA, June 2002. USENIX Association.

\bibitem{etsi-180001}
{Telecommunications and Internet converged Services and Protocols for Advanced
  Networking (TISPAN); NGN Release 1; Release definition}.
\newblock Technical Report 180 001 V1.1.1, ETSI, March 2006.

\end{thebibliography}

\end{document}